# Atomic-scale mechanism of enhanced electron-phonon coupling at the interface of MgB$_2$ thin film

Xiaowen Zhang[1,4,#], Tiequan Xu[2,#], Ruochen Shi[1,4], Bo Han[1,4], Fachen Liu[3,4], Zhetong Liu[3,4], Xiaoyue Gao[1,4], Jinlong Du[4], Yue Wang[2]* and Peng Gao[1,3,4]*.

[1] International Center for Quantum Materials, School of Physics, Peking University, Beijing 100871, China;

[2] Applied Superconductivity Center and State Key Laboratory for Mesoscopic Physics, School of Physics, Peking University, Beijing 100871, China;

[3] Academy for Advanced Interdisciplinary Studies, Peking University, Beijing 100871, China;

[4] Electron Microscopy Laboratory, School of Physics, Peking University, Beijing 100871, China;

[#] X.W. Zhang and T.Q. Xu contributed equally to this work.

* E-mail: yue.wang@pku.edu.cn; pgao@pku.edu.cn.

**Abstract**

In this study, we explore the heterointerface of $MgB_2$ film on SiC substrate at atomic scale using electron microscopy and spectroscopy. We detect ~1 nm MgO between $MgB_2$ and SiC. Atomic-level electron energy loss spectra (EELS) show $MgB_2$-$E_{2g}$ mode splitting and softening near the $MgB_2$/MgO interface. Orbital-resolved core-level EELS link the phonon softening to in-plane boron-atom electron states' changes. *Ab initio* calculations confirm this softening enhances electron-phonon coupling at the interface. Our findings highlight interface engineering's potential for superconductivity enhancement.

Magnesium diboride ($MgB_2$) stands out as a compelling superconducting material with one of the highest known transition temperature $T_c \approx 39$ K [1] among the intermetallic compounds, drawing considerable attention to $MgB_2$ thin films due to not only their significance in fundamental studies [2-5] but also their vast potential in diverse technological applications[6-9]. Of particular interest is the realm of $MgB_2$ ultra-thin films, offering an intriguing platform for the development of highly sensitive superconducting detectors, such as hot-electron bolometers [10] and kinetic inductance detectors [11], where achieving a higher $T_c$ is of paramount importance. Despite various methods have been explored to control $T_c$ of $MgB_2$ films, including strain engineering [12], doping [13], and isotopic substitution [14], the effectiveness of strain engineering is often constrained in ultra-thin films as their thickness [12], and doping typically leads to a decrease in $T_c$ [13].

For ultra-thin films, the role of the interface assumes critical significance [15-17], serving as boundary condition for electron wave functions. Previous investigation into the influence of the interface on the quantity of $MgB_2$ ultra-thin films has unveiled the profound impact of the substrate-surface termination on $MgB_2$ thin film quality [18]. Recent findings indicate that the interface can substantially enhance electron-phonon coupling caused by the strong interaction between Fuchs-Kliewer phonons of substrate and electrons [19-21], suggesting the possibility of managing $T_c$ of $MgB_2$ thin films through interface engineering. The underlying mechanism appears to hinge on the broken symmetry at the interface between $MgB_2$ thin film and the substrate, which alters phonon states and/or electron states [22-27]. Consequently, unraveling the relationship between the atomic structure, phonon states and electron states at the interface is not only pivotal for comprehending interface effects, but crucial for improving superconductivity performance via interface engineering, which motivates the present study.

However, the experimental characterization of the interface is challenging for the traditional methods like angle-resolved photoemission spectroscopy and Raman spectroscopy, which are usually limited in spatial resolution, hindering probing changes in electron and/or phonon behavior localized at the interface. Recent advances in scanning transmission electron microscopy - electron energy loss spectroscopy (STEM-EELS) with spherical aberration corrector and monochromator enables the imaging and spectroscopic characterization of materials simultaneously with atomic-scale resolution [28-30]. Besides, the wide-spectral range of STEM-EELS affords the unique capability to concurrently probe the electronic structures and phonon modes and further correlate them with specific local atom arrangements, creating unprecedented opportunities to explore the influence of the interface between $MgB_2$ films and substrates [31].

In this study, we use atomically resolved STEM-EELS in conjunction with density functional theory (DFT) to probe changes in phonon and electron states at the interface between $MgB_2$ film and SiC substrate, and reveal their relationship from the view of electronic band structure. We fabricate $MgB_2$ thin film on 6H-SiC substrate, but find that a thin MgO layer between $MgB_2$ and SiC exists, forming the $MgB_2$/MgO interface, at which the $E_{2g}$ phonon of $MgB_2$ softens, while the ratio $\pi^*/\sigma^*$ excitation of boron atom increases. The phonon softening corresponding to stronger electron-phonon coupling, holds the promise of superconductivity enhancement. The underlying mechanism driving this phonon softening lies in the difference in electronegativity between boron atoms and oxygen atoms, which causes electron transfer at the interface. These findings provide us new insights for tuning the superconductivity by using the interface engineering, which have implications for the design of ultra-thin $MgB_2$ thin-film devices.

The $MgB_2$ film was grown on the [001]-oriented 6H-SiC substrate using the hybrid physical-chemical vapor deposition (HPCVD) method. The X-ray diffraction pattern (Fig. S1(a)) reveals the film with its c-axis oriented normal to the substrate surface. Fig. 1(a) shows an atomically resolved high-angle annular dark-field (HAADF) image of the interface between $MgB_2$ thin film and 6H-SiC substrate with viewing along $MgB_2$ [11$\bar{2}$0] zone axis. The core-level EEL spectra were also recorded in Fig. S1(c) to determine the elemental distribution at the interface. Based on the HAADF image and EEL spectra, we find that a thin MgO layer is formed between $MgB_2$ and 6H-SiC which is consistent with previous research [18]. Previous study suggested that the 'native oxide' in the SiC substrate may account for the formation of MgO layer, and proposed that the interface could influence the growth habit of MgB$_2$ films [18]. The [111] zone axis of MgO is perpendicular to the interface. Fig. 1(b) displays the temperature dependence of the electrical resistance of this MgB$_2$ film under zero magnetic field, with onset and zero-resistance temperatures of 40.5 K and 40.2 K, respectively, which are closely align with those reported in prior research [32].

To investigate the effects of the interface on the MgB$_2$ phonon modes, we measured the phonon spectra from MgB$_2$ to MgO and their interface using atomic-resolution STEM-EELS. Line profile of EEL spectra is shown in Fig. 2(a), revealing distinct atomic-column contrast, which arises from the fact that magnesium atom mainly contributes to lower-energy acoustic vibrations, while boron atom mainly contributes to high-energy vibrations (details in Fig. S2). These observed atomic-scale differences in EEL spectra also indicate that the STEM-EELS allows to detect the phonon changes near the interface. Fig. 2(b) provides detailed changes in EEL spectra at boron atomic columns, ranging from bulk (labeled as Bn, where n denotes the number of boron atomic layers away from the interface) to the interface (labeled as IB). Within these spectra, three primary spectral features can be observed from B7 to B1, labeled as P1, P2, and P3, respectively. These features were identified by Gaussian peak fitting and are depicted as gray, orange, and cyan shades. By comparing these findings with DFT calculations (Fig. S3), we infer that P1 corresponds to the acoustic phonon contribution, P2 primarily represents the $E_{2g}$ mode, and P3 arises from the transverse optical (TO) branch phonon located at the edge of the Brillouin zone. As approaching the interface (IB), P3 broadens, while P2 (~66 meV) splits into two branches with substantial softening, i.e., P2' (~60 meV) and P2'' (~49 meV).

To understand the underlying cause of these changes, we performed DFT calculations on the phonon spectra of the interface model, and the projected phonon density of states (PPDOS) exhibits a similar feature (details in Fig. S4). By examining the eigenvectors of calculated phonon modes in Fig. 2(c), the splitting and softening of P2 result from in-plane anti-phase vibrations of boron atoms at the interface, which resembles the $E_{2g}$ mode observed in the bulk. The $E_{2g}$ phonon mode in bulk MgB$_2$ is doubly degenerate at Γ, involving in-plane, anti-phase stretching and hexagon-distorting displacements of the boron atoms [33]. Furthermore, utilizing the density-functional perturbation theory (DFPT) method, we calculated the electron-phonon coupling strength, and found that the softening-akin-$E_{2g}$ mode at the interface exhibits stronger coupling (Fig. S5), i.e., the electron-phonon coupling constant at the Γ point increases from ~1.1 in the bulk to ~1.5 at the interface. In-plane $E_{2g}$ phonon mode plays a predominant role in electron-phonon coupling [34], for $MgB_2$, so we believe such interface-induced enhancement of electron-phonon couplings hold potential for improving the superconductivity in ultra-thin films of $MgB_2$.

To thoroughly comprehend the origin of $E_{2g}$ mode softening at the interface, we analyzed the differences in chemical bonding between bulk and interface. Fundamentally, due to the interfacial Mg-O bonding, the electron states of interfacial boron atoms are modified, impacting the vibration behavior of the interfacial boron atoms. Similar to graphene [35], the chemical bonding of boron atoms in $MgB_2$ can be described as in-B-plane ($\sigma$) and orthogonal out-of-B-plane ($\pi$) covalent bonds [36]. The EELS near edge structure of the B-K edge can provide information for insight into the excitation of core states to probe in-B-plane $\sigma^*$ (1s → $2p_{x,y}$) orbitals and out-of-B-plane $\pi^*$ (1s → $2p_z$) orbitals. For this reason, we measured the core-level EEL spectra with atomic-resolution, focusing on the boron atoms since the $E_{2g}$ mode is mainly contributed by their in-plane vibrations. Fig. 3(a) displays the line profile of atomic-resolution core-level EEL spectra along with the corresponding HAADF image. The EEL spectra plotted in Fig. 3(b) show an energy difference of 3.5 eV between $\sigma^*$ orbital excitation and $\pi^*$ orbital excitation, which closely matches the corresponding values from DFT calculations (Fig. 3(c)). The energy difference between $\sigma^*$ and $\pi^*$ is extracted using second differential of EEL spectra (see details in Fig. S6). Notably, although the intensities of both $\pi^*$ and $\sigma^*$ signals diminish approaching interface, the intensity of the ratio $\pi^*/\sigma^*$ increases, as shown in Fig. 3(d) and Fig. 3(e). From the DFT calculations, we found that the elevation of $\sigma^*$ electron state's energy level at the interface plays a significant role. Fig. S7 illustrates that the electron states projected onto boron's atomic orbitals above the Fermi energy at the Γ point experience an energy rise at the interface compared to the bulk states. The higher energy of the $\sigma^*$ electron state implies less occupation of electrons according to Fermi-Dirac distribution, and thus weaker strength of the in-plane B-B bonding, consequently contributing to a softer phonon mode. Fig. 3(f) depicts the relationship between the in-plane electron charge density of boron atoms and the $E_{2g}$ mode. The frequency of $E_{2g}$ is decided by in-plane bonding of boron atoms, governed by electron states.

Over an extended period, the atomic structure of the interface between $MgB_2$ film and substrate has remained enigmatic, with its consequential influence largely uncharted. Our study simultaneously determined the atomic structure, phonon modes and electron states at the interface. These results unveiled an unexpected revelation—the presence of MgO at the $MgB_2$/SiC interface, diverging significantly from the anticipated "$MgB_2$/SiC" interface. This unanticipated presence of MgO exerted a profound influence, initiating unanticipated chemical bonding, then phonon behavior. Moreover, such an unexpected $MgB_2$/MgO interface shows the enhanced electron-phonon coupling of $MgB_2$ due to the unique electron states at the interface. These findings firmly establish the precise relationship between the atomic structure, phonon modes and electron states at the interface, which may help us to understand the past experiments that the $MgB_2$ thin film on SiC substrate usually has higher superconducting transition temperature than bulk from the view of interface enhancement [32].

Furthermore, the underlying mechanism for such an interface phonon softening is also revealed, i.e., the unsaturation of bonding among boron atoms at the interface, stemming from the heightened electronegativity of oxygen in comparison to boron, leading to weaker vibration along the in-plane anti-phase direction ($E_{2g}$ mode). Therefore, our results carry significant implications for the practical implementation of ultra-thin $MgB_2$ films in the device applications that require higher $T_c$, as, for example, higher $T_c$ means lower cooling cost in bolometers [37]. To optimize the superconductivity, previous strategies mainly involve strain engineering and doping, while our study introduces a novel avenue: the modulation of local electron states through interface engineering, which may be new avenue to tune the strength of electron-phonon coupling in $MgB_2$.

In conclusion, our study revealed the significant influence of the interface on phonon behavior in MgB$_2$ films, and found this behavior originates from the changes of in-plane electron states of boron atoms owing to unique atomic arrangement of the interface. These findings underscore the importance of interface engineering to optimize the superconducting properties, particularly in ultra-thin film devices. STEM-EELS, enabling simultaneous determination of the atomic structure, phonon states and electron states, offers a powerful approach for investigating the properties of interfaces, further facilitating the design and development of advanced materials and devices.

This work was supported This work was supported by the National Key R&D Program of China (2021YFA1400500) and the National Natural Science Foundation of China (52125307, 11974023, 52021006). We acknowledge Electron Microscopy Laboratory of Peking University for the use of electron microscopes, and the High-performance Computing Platform of Peking University for providing computational resources.

Figures

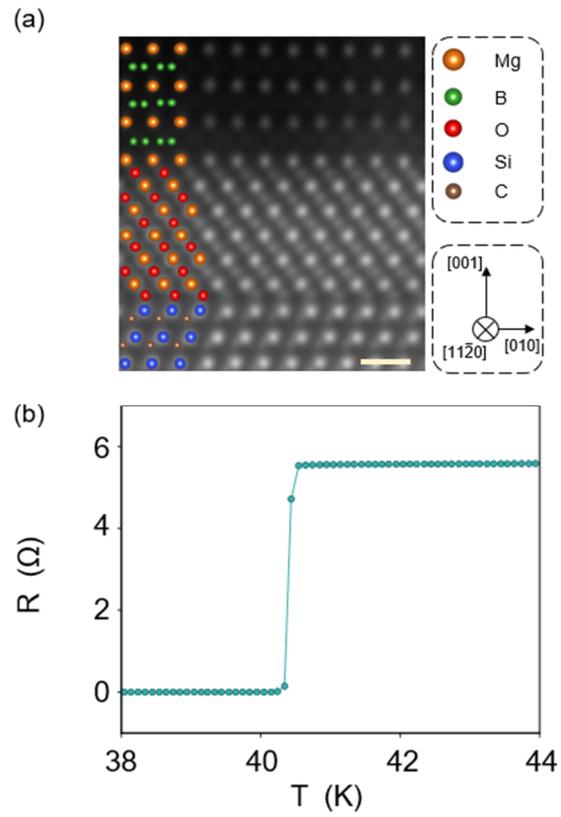

FIG. 1. Atomic structure and transport property of MgB$_2$ thin film on 6H-SiC substrate. (a) Atomically resolved HAADF image and corresponding atomic arrangements at the region between film and substrate. Scale bar, 0.5 nm. (b) Superconducting transition of the film in resistance measurement under zero magnetic field.

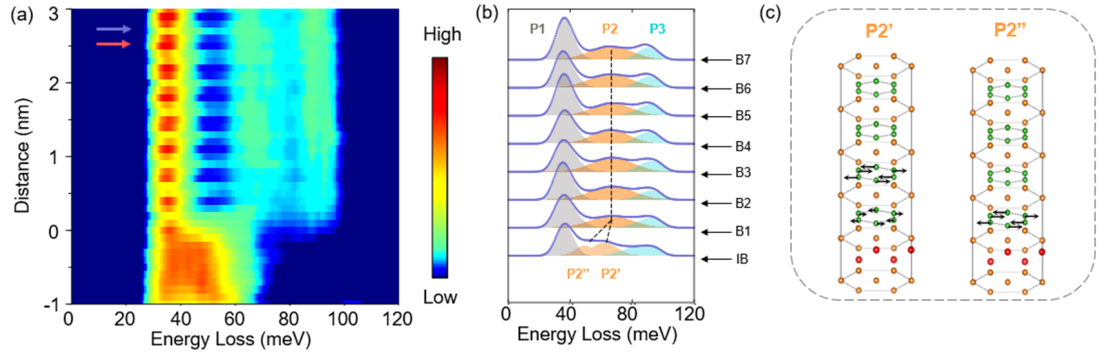

FIG. 2. Atomically resolved phonon spectra at the interface. (a) Line profile of EEL spectra. The red arrow represents the Mg atomic column layer, while the purple arrow displays the B atomic column layer. The 0 nm labels the location of the interface between $MgB_2$ and MgO. (b) Fitted phonon peaks of different branches approaching the interface. Bn and IB label the bulk-boron-column and interface-boron-column, respectively. (c) The eigenvectors of the modes that mainly contributes to P2' and P2'', individually.

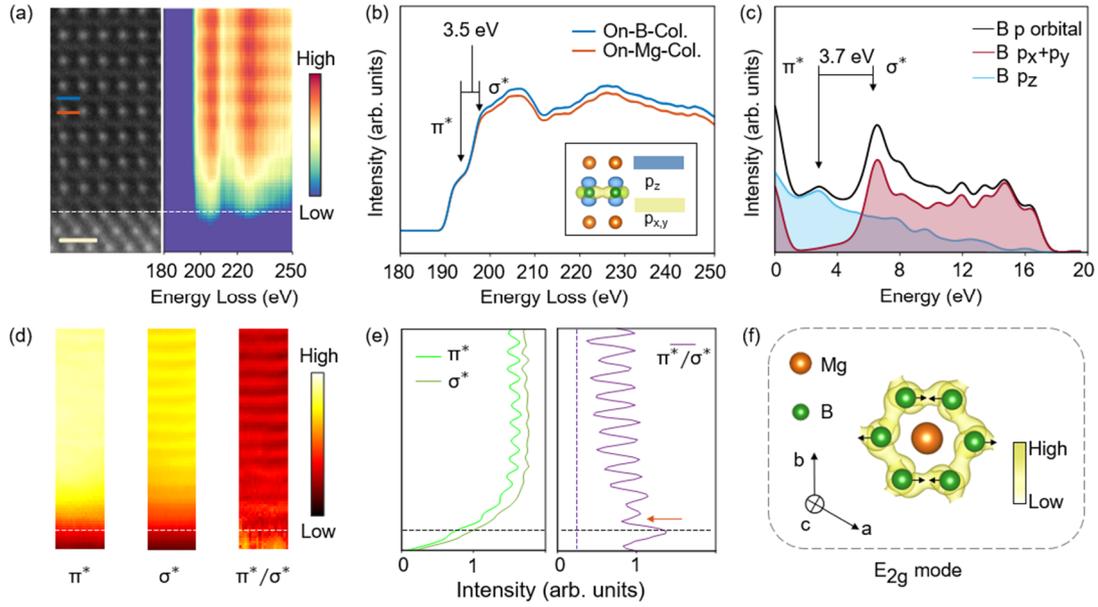

FIG. 3. Atomically resolved electronic structures at the interface. (a) A HAADF image and the corresponding core-level EEL spectra. The white horizontal dashed line indicates the interface of $MgB_2$/MgO. (b) B-K edge spectra corresponding to the probe positioned on-B-column (solid blue line) and on-Mg-column (solid orange line), as indicated in (a). Two arrows indicate the $\pi^*$ and $\sigma^*$ signal, respectively. The charge density of in-B-plane p orbitals and out-of-B-plane p orbitals are shown in the bottom-right location in (b), labeled with yellow and blue shade, correspondingly. (c) Projected density of states in $MgB_2$. The z axis corresponds to the crystallographic c axis of $MgB_2$. Two arrows label the $\pi^*$ and $\sigma^*$ signal, as (b). (d) $\pi^*$, $\sigma^*$ and $\pi^*/\sigma^*$ maps. The white dashed lines label the interface. (e) Corresponding profiles integrated along the direction paralleled to the interface. The black horizontal dashed lines indicate the interface, while the vertical one is for the eye's guide. The red arrow indicates the location of the boron column at the interface. (f) The relation between the charge density of in-B-plane orbitals and the strength of $E_{2g}$ mode in $MgB_2$.

# Supplemental Material for

# Atomic-scale mechanism of enhanced electron-phonon coupling at the interface of MgB$_2$ thin film

Xiaowen Zhang[1,4,#], Tiequan Xu[2,#], Ruochen Shi[1,4], Bo Han[1,4], Fachen Liu[3,4], Zhetong Liu[3,4], Xiaoyue Gao[1,4], Jinlong Du[4], Yue Wang[2]* and Peng Gao[1,3,4]*.

[1] *International Center for Quantum Materials, School of Physics, Peking University, Beijing 100871, China;*

[2] *Applied Superconductivity Center and State Key Laboratory for Mesoscopic Physics, School of Physics, Peking University, Beijing 100871, China;*

[3] *Academy for Advanced Interdisciplinary Studies, Peking University, Beijing 100871, China;*

[4] *Electron Microscopy Laboratory, School of Physics, Peking University, Beijing 100871, China;*

[#] X.W. Zhang and T.Q. Xu contributed equally to this work.

* E-mail: yue.wang@pku.edu.cn; pgao@pku.edu.cn.

1. Methods

**Sample Preparation.** The growth of $MgB_2$ film on 6H-SiC substrate is using the HPCVD method, whose details have been described previously [1]. TEM samples were prepared by the focused ion beam (FIB) technique (Hitachi FB2200 FIB) with a gallium ion source. To clean the damaged layer of the TEM samples induced by the ion radiation of FIB, argon ion-milling with an accelerating voltage of 0.5 kV was performed using a precision ion polishing system (Model 691, Gatan). The film thickness in our region of interest is 20–30 nm.

**EELS Acquiring.** The vibrational spectra were acquired at a Nion U-HERMES200 electron microscope equipped with both the monochromator and the aberration corrector operated at 60 kV. The probe convergence semi-angle was 35 mrad, while the collection semi-angle was 25 mrad. The electron beam was moved off optical axis with 60 mrad for off-axis experiments to greatly reduce the contribution of the dipole scattering. The energy dispersion channel was set as 0.5 meV with 2,048 channels in total. The spatially resolved EEL spectra in Fig. 2(a) was originally recorded as a spectrum image, with single exposures of 1,200 ms per pixel. The acquired spectrum images are 2 nm × 8 nm with 0.1 nm per pixel in Fig. 2(a). As to the core-level EEL spectra in Fig. 3(a), it was recorded as spectrum image with 2 nm × 8 nm and 0.0625 nm per pixel. The HAADF images were recorded by an aberration-corrected FEI Titan Themis G2 operated at 300 kV.

**EELS Processing.** All acquired EEL spectra were processed using the Gatan Microscopy Suite and custom-written MATLAB code. The EEL spectra were first aligned by their normalized cross-correlation. Next, the block-matching and three-dimensional filtering (BM3D) algorithm were applied to remove gaussian noise.

For the vibrational spectra, the background arising from both the tail of ZLP and non-characteristic phonon losses was fitted using the modified Peason-VII function with two fitting windows and then subtracted to obtain the vibrational signal. The Lucy–Richardson algorithm was then employed to ameliorate the broadening effect caused by finite energy resolution, taking the elastic ZLP as the point spread function. But for the core-level EEL spectra, their ZLP were subtracted using power law function. The spectra were summed along the direction parallel to the interface to obtain a line-scan data with a good signal-to-noise ratio. The vibrational spectra were fitted using a simple Gaussian peaks-fitting model to extract the peak positions.

**DFT calculations.** Density functional theory calculations were performed using Quantum ESPRESSO [2] with the Perdew–Zunger exchange-correlation functional [3] and the Vanderbilt ultrasoft pseudopotential [4]. The kinetic energy cut-off was 50 Rydbergs (Ry) for wavefunctions and 500 Ry for charge density and potential. The interface model contains 4 unit-cell $MgB_2$ connected to 2 unit-cell MgO and then 1 unit-cell SiC (35 atoms in one hexagonal unit cell with cell parameters $a$ = 3.085 Å and $c$ = 50 Å). The small lattice mismatch among them was ignored. The structure was optimized until the residual force was below $10^{-4}$ Ry per Bohr on every atom. The dynamical matrices and force constants were obtained using DFPT. The phonon dispersion and PDOS was calculated by interpolating the dynamical matrix on a $3 \times 3 \times 1$ q-mesh. Electron–phonon coupling calculation for the interface model was performed with a $16 \times 16 \times 1$ dense mesh of k-points for the electron–phonon coefficients at the Fermi energy, as implemented in Quantum ESPRESSO. The DFT calculations of bulk $MgB_2$ were performed with the same functional and pseudopotential as the interface model. The kinetic energy cut-off was 50 Rydbergs (Ry) for wavefunctions and 500 Ry for charge density and potential. The structure was optimized until the residual force was below $10^{-4}$ Ry per Bohr on every atom. The dynamical matrices and force constants were obtained using DFPT. The phonon dispersion and PDOS was calculated by interpolating the dynamical matrix on a $4 \times 4 \times 4$ q-mesh.

## 2. Figures

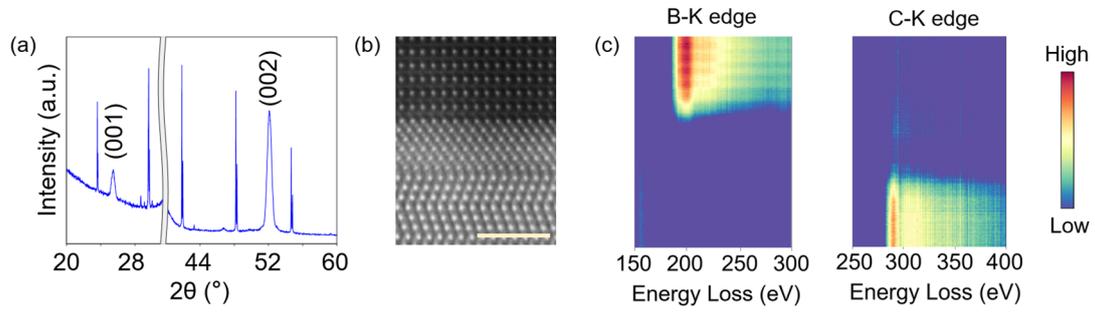

FIG. S1. Characterization of MgB$_2$ thin film on 6H-SiC substrate. (a) X-ray diffraction of the MgB$_2$ film grown on the 6H-SiC substrate. The peaks from MgB$_2$ (001) and (002) are indicated and other peaks are from the substrate. (b) Atomic HAADF images. Scale bar indicates 2 nm. (c) Left panel: Line profile of B-K edge spectra, corresponding to (b). Right panel: Line profile of C-K edge spectra, corresponding to (b).

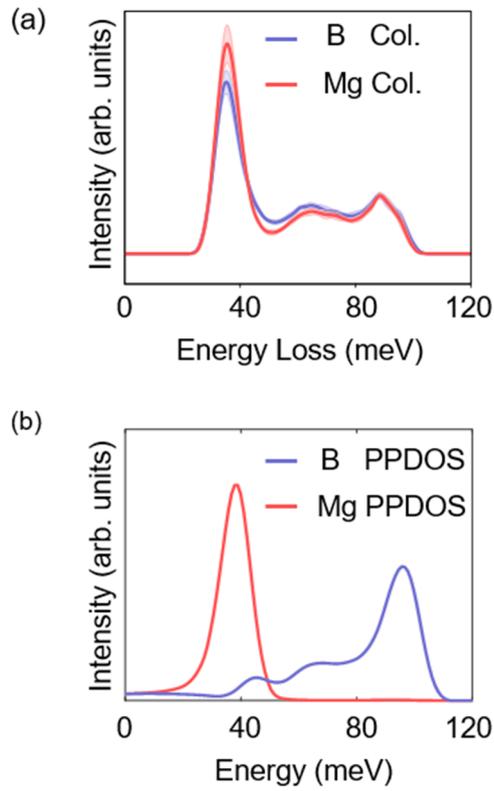

FIG. S2. EEL spectra and PPDOS of different atoms in MgB$_2$. (a) EEL spectra on different atomic columns. The purple solid line shows the EEL spectra on B atomic columns, while the red one displays spectra on Mg. (b) PPDOS of different kinds of atom. The purple solid line shows PPDOS of B atoms, while the red one displays PPDOS of Mg.

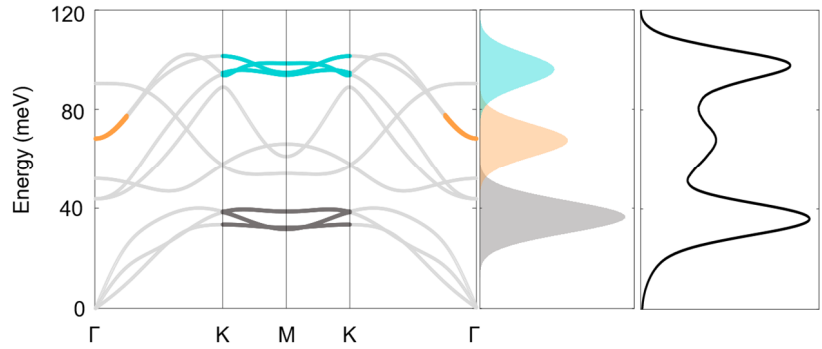

FIG. S3. Phonon dispersion of MgB$_2$, along ΓKMKΓ direction. The gray shade is contributed by the acoustic phonon, and the orange shade represents the contribution of near Γ point, and the cyan shade represents the TO phonon at the edge of Brillouin zone. The origins of these shade are labeled in the phonon dispersion, with their corresponding color. The total PPDOS is shown in the right part, plotted with black solid line.

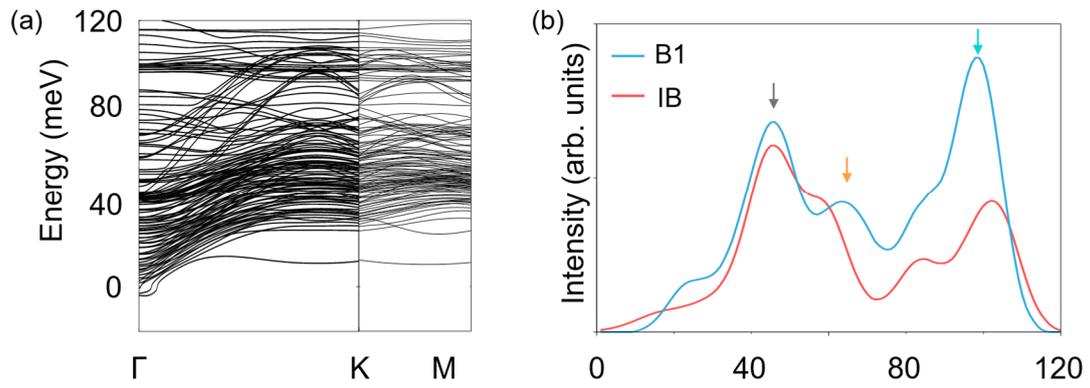

FIG. S4. Phonon spectra and electron-phonon coupling in the interface model. (a) Phonon dispersion in the interface model along ΓKM. The small imaginary frequency means that the structure is dynamically stable. (b) The comparison of PPDOS between the interface B atoms (labeled by IB) and far from interface B atoms (labeled by B1). Gray, orange and cyan arrow mean peaks displayed in Fig. S3.

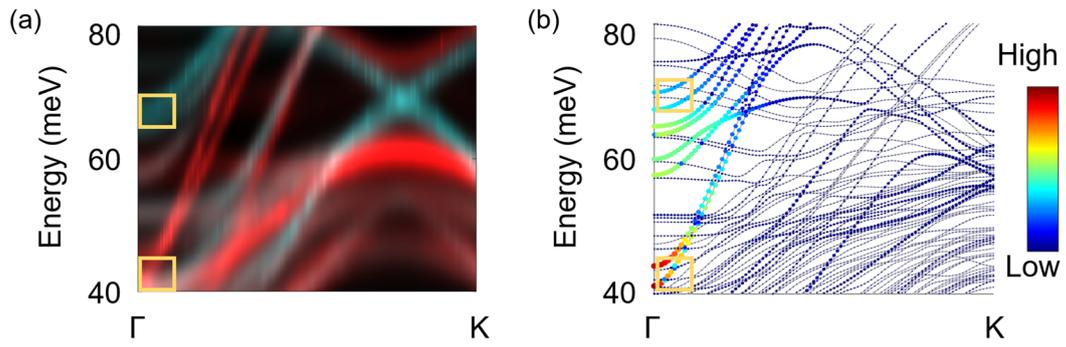

FIG. S5. (a) Projected phonon dispersion along ΓK, on the bulk boron atoms (cyan line) and the interface boron atoms (red line). (b) Electron-phonon coupling strength dispersion along ΓK. Golden solid boxes in (a) and (b) show the in-plane anti-phase vibrations of boron atoms in B1 (up) and IB (down), respectively.

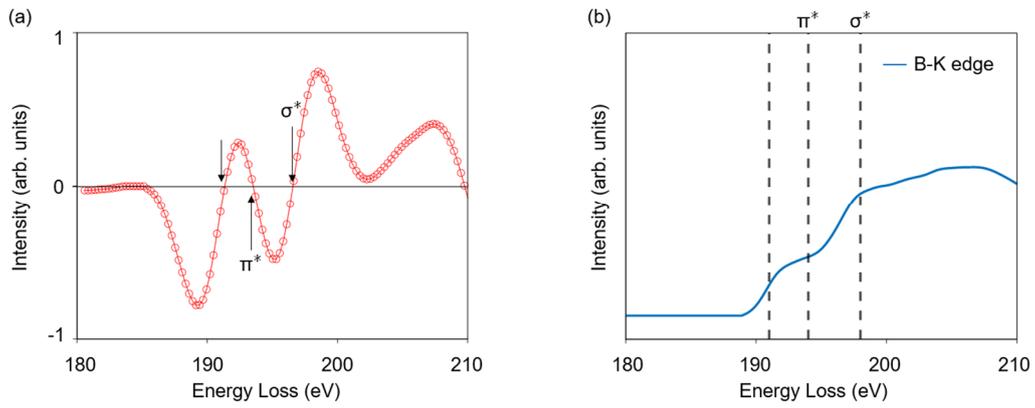

FIG. S6. Identification of $\pi^*$ and $\sigma^*$ signals in EEL spectra. (a) 2$^{nd}$ order derivate of B-K edge spectra. Black arrows indicate the zero-point, respectively. (b) B-K edge spectra. Black dashed vertical lines mean the zero-point indicated in (a).

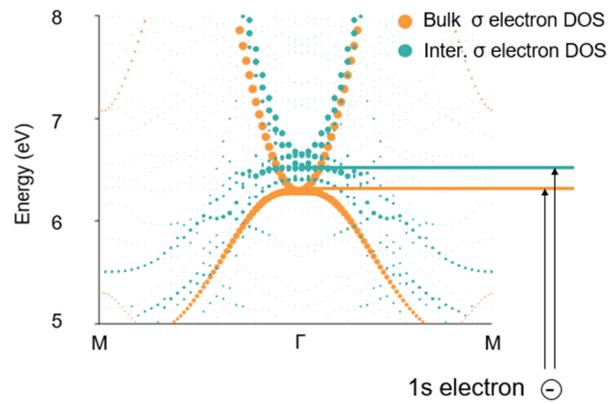

FIG. S7. Projected electron state dispersion of bulk-boron atoms and interface-boron atoms. The size of colored solid circles means the contribution to projected density of state. The solid black arrows mean the process of the excitation of $\sigma^*$ in bulk-boron atoms and interface-boron atoms, respectively. The orange and green line indicate the energy level in $\Gamma$ point, respectively.